\documentclass[conference]{article}
\usepackage{amsmath,amssymb,amsfonts}
\usepackage{algorithmic}
\usepackage{graphicx}
\usepackage{textcomp}
\usepackage{xcolor}

\def\BibTeX{{\rm B\kern-.05em{\sc i\kern-.025em b}\kern-.08em
    \kern-.1667em\lower.7ex\hbox{E}\kern-.125emX}}
\begin{document}

\title{Web Application Attack Detection using Deep Learning}
 
\author{Tikam Alma and Manik Lal Das\vspace{2 mm}\\
DA-IICT\\
Gandhinagar, India\\
Email: \texttt{\{201601030, maniklal\_das\}@daiict.ac.in}}

\date{}

\maketitle

\begin{abstract}
Modern web applications are dominated by HTTP/HTTPS messages that
consist of one or more headers, where most of the exploits and
payloads can be injected by attackers. According to the OWASP, the
80 percent of the web attacks are done through HTTP/HTTPS requests
queries. In this paper, we present a deep learning based web
application attacks detection model. The model uses auto-encoder
that can learn from the sequences of word and weight each word or
character according to them. The classification engine is trained
on ECML-KDD dataset for classification of anomaly queries with
respect to specific attack type. The proposed web application
detection engine is trained with anomaly and benign web queries to
achieve the accuracy of receiver operating characteristic curve of
1. The experimental results show that the proposed model can
detect web applications attack successfully with low false
positive rate.\vspace{1 mm}

\end{abstract}

\section{Introduction}
Web application attacks are found one of the most targets by
attackers. Symantec Internet security \cite{istr2019} reported an
interesting statistics that 1 in 10 URLs are identified as being
malicious. Web applications typically use the HTTP/HTTPS protocols
supported by other backend and frontend interfaces. According to
Imperva Web Application Vulnerability Report \cite{istr2019}, high
severity attacks are injection attacks, which are being exploited
through injecting payloads in HTTP/HTTPS web queries by using GET,
POST and PUT methods. CSRF (Cross Site Request Forgery) attack,
SQL injection attack, XSS (Cross Site Scripts) attacks, and widely
used vulnerable JS libraries, which account for 51 percent, 27
percent, 33 percent and 36 percent, respectively. This paper
focuses on the most frequent types of web-based injection attacks,
which includes SQL injection, XSS (Cross Site Script), RFI (Remote
File Inclusion), XXE (XML External Entity), CSRF (Cross Site
Request Forgery), and SSRF (Server Side Request Forgery).\vspace{1 mm}\\
Network Intrusion Detection System (NIDS) monitors the network
traffic in web applications. Web IDS acts as intermediate between
web application and users, as it analyzes web traffics to detect
any anomaly or malicious activity \cite{ying2017}. Generally,
there are two types of detection approaches: anomaly-based
detection and signature-based detection. The signature-based IDS
system uses a signature concept, more like antivirus detects the
virus, when the antivirus database has that specific kind of virus
signature. If attackers create new virus, the antivirus is of no
use if the signature/pattern is not present. Anomaly-based
detection is based on detection unique behavior pattern
recognition or any activity that differs from previous data or
information is fed. When comparing signature-based detection
method with anomaly-based detection method, the performance of
anomaly based detection found high to detect the unknown attacks,
but it comes with a cost that it has problem with false positive
alarm rates. After detection of an anomaly it is stored in the
database it becomes a ``signature'', and furthermore, there are
two detection methods which come under anomaly detection which is
adaptive detection and constant detection \cite{lstm}. An adaptive
detection algorithm analyzes the network traffic of port 80 of web
network that is HTTP traffic, which continuously gets the input of
traffic and analyzes in a timely manner, while constant based
detection method analyzes stores incoming traffic or use to
analyze the logs of collected traffic.\vspace{1 mm}\\
The conventional patching approach to mitigating most network
layer vulnerabilities does not work well in web application
vulnerabilities such as SQLi, RCE, or XSS. The reason behind all
these attacks is that modern web applications are poorly designed
with insecure coding. One can follow the OWASP's secure coding
guidelines to prevent most of the attacks. Adaptive detection
model is effective to detect anomalies and classify them which
type of attack it is, so that the developer at backend can fix
that patch or prevent it (e.g., Django uses CSRF Tokens in the
framework to prevent CSRF) attacks which account for 51 percent of
the web attack. At the same time, the model should learn the
patterns over time to detect unknown web attacks and identify
which type of attack vectors are being exploited. Anomaly based
detection approaches \cite{ying2017} usually rely on an adaptive
model to identify anomalous web requests, but with a high degree
of false positive rate. In this paper, we come up with a solution
to handle false positive where IDS monitor a system based on their
behavior pattern. There are several reasons why a conventional IDS
or web application firewall does not work, as follows:

\begin{itemize}
\item Limited Dataset: To collect and capture a large amount of
anomalous data, one has to set-up a system or an automated system
that captures the attack requests and classify them whether they
are anomalous or normal requests. This is more like a
attack-defense simulation system, but smart enough to classify,
that does not need to be labeled manually and it could save lots
of time.\vspace{2 mm}

\item High False Positive: Conventional system uses unsupervised
learning algorithms such as PCA \cite{nn1} and SVM \cite{nn2} to
detect web attacks, these approaches require manual selection of
attack specific features \cite{yp}. These conventional methods may
achieve acceptable performance, but they face high false positive
rates.\vspace{2 mm}

\item Labeled Dataset: Conventional IDS uses rule-based or
conditional strategies or supervised algorithms like support
vector machines or decision trees to classify normal traffic
requests from attack requests, which requires large database to
get the accurate results \cite{yp}.
\end{itemize}
In this paper, we present a web application attacks detection
model, SWAD, based on deep learning technique that detects web
application attacks autonomously in real-time. The model uses
auto-encoder that can learn from the sequences of word and weight
each word or character according to them. The classification
engine is trained on ECML-KDD dataset for classification of
anomaly queries with respect to specific attack type. We have
implemented the model on sequence to sequence model, which
consists of encoder and decoder, that sets its target values equal
to its input values. The proposed SWAD model first uses 40,000 web
requests, both anomaly and benign nature, for training and then
20,000 anomaly web requests and responses for training the model.
The experimental results of the proposed model show that it can
detect web applications attack with true positive rate is 1 and
low false positive rate.\vspace{2 mm}\\
The paper is organized as follows: Section II summarizes the
background and related works. Section III describes the system
design. Section IV evaluates the performance of the proposed
model. Section V concludes the paper.

\section{Background and Related work}

\subsection{Deep Learning for Web Attack Detection}
There are two categories of Machine Learning approaches for
detecting web attacks: unsupervised and supervised learning.
Supervised learning is the learning approach feeds mapped labeled
data which then outputs the expected data, which is simply mapping
of input functions to dataset and expecting new input with learned
labels at the output. For the classification of data, the most
common algorithm is supervised learning, which is used to learn
the machine learning model to train and identify the data using
the labels which are mapped. The concept of the algorithm is to
learn the mapping function of a given input to the output, where
the output is defined by variable $Y$ and input is defined by
variable $x$.

\[ Y = f(X) \]

If web attacks labeled dataset is trained using supervised
algorithms such as SVM (Support Vector Machine) \cite{svm} and
Naive Bayes \cite{nb}, then it classifies anomalous to normal web
attack requests. However, the model cannot handle new types of
attack requests and it requires a large amount of labeled
dataset.\vspace{1 mm}\\
Unsupervised learning is used mainly with unlabeled dataset. The
model supported by this learning finds patterns from previous
sequence or dataset and identifies or predicts the next one. For
exploratory analysis, the unsupervised learning method is used to
automate the identification pattern of data structures. Using an
unsupervised method, one can reduce the dimensions used to
represent data for fewer columns and features. To sort the
eigenvectors, Principal Component Analysis (PCA) \cite{pca} is
used to compute the eigenvectors of the co-variance matrix that is
``principal axes''. To get the dimensionality reduction, the
centered data were projected into principal axes. Principal
component (PC) scores are a group of score that are obtained from
PCA. To create equal number of new imaginary variables or
principle components the relationship between a batch or group of
those PC scores are analyzed. The optimized and maximally
correlated with all of the original group of variables is the
first created imaginary variables, then the next created imaginary
variable is less correlated and next is lesser than the previous
and it goes on until the point when the principal components
scores predicts any variable from the first created group.

PCA Reconstruction = PC Score $\times$ EigenVectors(t) +
Mean \vspace{1 mm}\\
The condition of perfect reconstruction of the data or input and
the there will be no dimentionality reduction is when all the $p$
eigenvectors are used and $VV^t$ is the identity matrix. When
using large dataset features, whether it is image, text or video
data, one cannot use any machine-learning algorithms directly. In
order to reduce the training time, prepossessing steps are
required to clean the dataset. It is noted that PCA is restricted
to a linear map. Autoencoders \cite{acd} can have non linear
encoder/decoders. A single layer autoencoder with linear function
is nearly equivalent to PCA. We use sequence-to-sequence
autoencoder in our proposed detection model.

\begin{figure}[h!]
\centering
  \includegraphics[width=7cm, height=4cm]{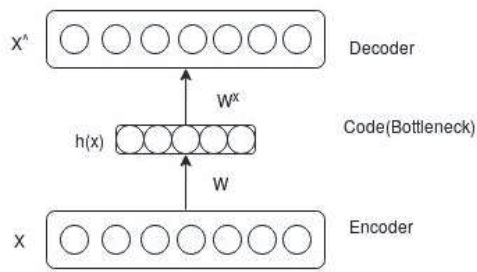}
  \caption{Autoencoder basic architecture}
\end{figure}

The condition to make the autoencoder equivalent to principal
component analysis is that if normalized inputs are used with the
linear decoder, linear encoder and square error loss function,
thenautoencoders are not restricted to linear maps. The proposed
model is optimized and trained to minimize and reduce the loss
between the input and the output layer. We have used non-linear
functions with encoders to get more accuracy when reconstruction
of data is processing. The activation functions used in
autoencoders are ReLu and sigmoid, which are non-linear in nature.

\begin{equation}
    \Phi : \chi \rightarrow F
\end{equation}

\begin{equation}
        \Psi : F \rightarrow \chi
\end{equation}

\begin{equation}
    \Phi , \Psi = arg min_\Phi,_\Psi || X = (\Phi * \Psi) X || ^2
\end{equation}
The encoder function, denoted by \( \Phi \), maps the original
data X to a latent space F. The decoder function, denoted  by \(
\Psi \), maps the latent space F to the output. We basically
recreate the original image after some generalized non-linear
compression. The encoding network can be represented by the
standard neural network function passed through an activation
function, where $z$ is the hidden dimension. The output works the
same as the input.

\begin{equation}
    Z = \sigma(W_x + b)
\end{equation}
With slight different weight, bias and activation function, the
output function or the decoder network is represented in the same
way.

\begin{equation}
     X^{'} = \sigma^{'}(W^{'}z + b^{'})
\end{equation}
To train the model for getting optimized results and the loss
function in the equation, the model is trained with
back-propagation method.

\begin{equation}
    L(x, x^{'}) = || x - x^{'} || = || x -{\sigma^{'}}({W^{'}}(\sigma(Wx+b))+{b^{'}})||^2
\end{equation}
To reconstruct the input data or input characters, the
autoencoders select the encoder and decoder function for
optimization, so that it requires the minimal information to
encode the input data for reconstructing the output.

\section{The Proposed Model}\label{AA}
The proposed detection engine uses an autoencoder model based on
sequence to sequence architecture that is made up of LSTM (Long
Short Term Memory) \cite{lstm} cells. LSTM networks are complex
neural networks that are used to train ordered sequences of inputs
to remember it and re-create it. The proposed model devises the
LSTM neural network model, which feeds sequenced inputs. After
completing the reading input processes, the output is given by an
internal learned representation of the fed input sequences as a
fixed-length vector. Then, output vector is fed inputs that
interprets the input sequence by sequence at each step and the
output is generated.

\begin{figure*}[h!]
  \includegraphics[width=\textwidth, height=12cm]{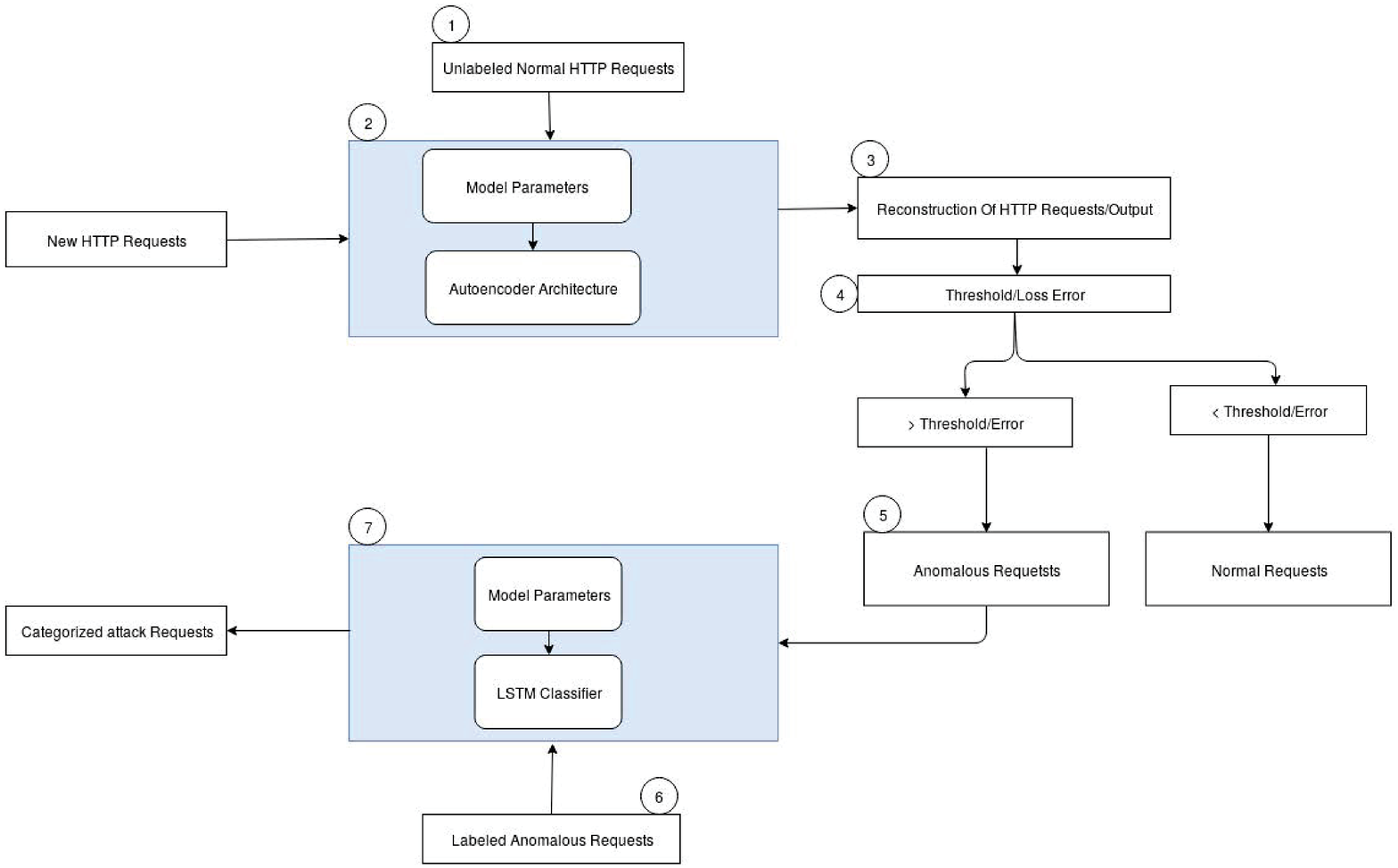}
  \caption{The Proposed System Architecture}
\end{figure*}
The proposed detection and classification model works in
synchronization as follows:

\begin{enumerate}
\item For the training purpose, large amounts of unlabeled normal
HTTP requests are collected from open-source Vulnbank
organization, which contains 40k normal HTTP (GET,POST and PUT)
methods requests.\vspace{2 mm}

\item For the auto-encoder's (Encoder-Decoder) architecture, the
hyper-parameters are trained by setting the problem as a grid
search problem. Each hyper-parameter combination requires training
the neuron weights for the hidden layer(s), which results in
increasing computational complexity with an increase in the number
of layers and number of nodes within each layer. To deal with
these critical parameters and training issues, stacked
auto-encoder concepts have been proposed that trains each layer
separately to get pre-trained weights. Then the model is
fine-tuned using the obtained weights. This approach significantly
improves the training performance over the conventional mode of
training. For implementation of the proposed model, we consider
the following parameters.\vspace{1 mm}\\

Batch Size = 128

Embed Size = 64

Hidden Size = 64

Number of Layers = 2

Dropout Rate = 0.7\\

\item Reconstruction of requests are done by the decoder
\(X^{'}=\sigma^{'}(W^{'}+b^{'})\), which perfectly reconstructs
the given input and evaluates loss function and accuracy.\vspace{2
mm}

\item When a new requests is given as input to the trained
autoencoder, it decodes and encodes the requests vector and
calculates the reconstruction or loss error. If loss error is
larger than the learned threshold \(\theta\), it categorizes as
anomalous requests. If loss error is smaller than \(\theta\), it
categorizes as normal requests.\vspace{2 mm}

\item After categorizing requests into \textit{normal} and
\textit{anomalous} requests, normal requests are sent to the
database for retraining or re-learning, so that over time the
detection model learns new type of requests patterns. Anomalous
requests are sent to the classification model which further
categorizes the anomalous requests into which type of attack it
was exploited through requests like SQLi, XSS or CSRF.\vspace{2
mm}

\item The classification model is trained on larger number of
labeled attack vectors HTTP requests. It contains 7 class of
attacks which are Os-Commanding, PathTraversal, SQLi,
X-PathInjection, LDAPInjection, SSI, and XSS.
\end{enumerate}
We use LSTM layers to train the classification model and fine-tune
the model with hyperparametrs. Every LSTM layer is accompanied by
a dropout layer, which helps to prevent over-fitting by ignoring
randomly selected neurons during training, and hence, reduces the
sensitivity to the specific weights of individual neurons.

\begin{figure}[h!]
  \includegraphics[width=7cm,height=4cm]{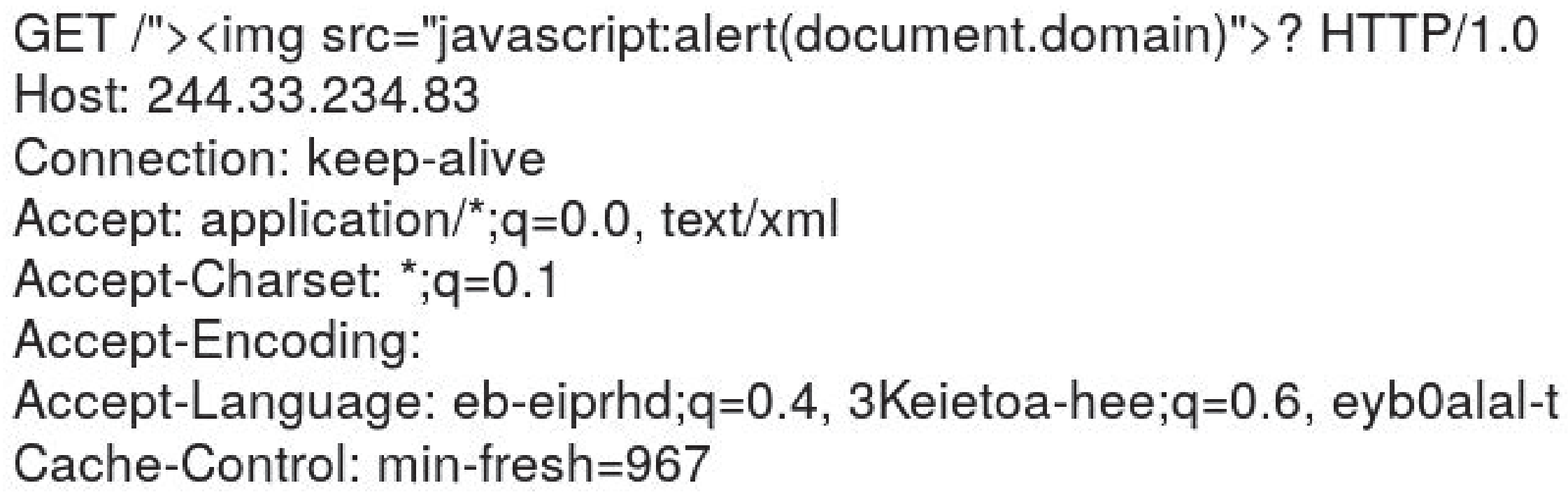}
  \caption{HTTP Requests with XSS attack Vector}
\end{figure}
The image in Figure-3 is the raw anomaly HTTP requests with XSS
attack vector. Tn data pre-processing step, the raw HTTP data is
converted to a single string and parsed as input to the LSTM cell,
which is the passed to the training phase to train the model.

\section{Experimental Results and Evaluation}
We have experimented the proposed model with 40,000 web requests
followed by 20,000 anomaly web requests and responses. The
classification engine is trained on ECML-KDD dataset for
classification of anomaly queries with respect to specific attack
type. We have evaluated the proposed model on ROC curve. An ROC
curve is a graph showing the performance of a classification model
at all classification thresholds. The ROC curve plots two
parameters - true positive rate and false positive rate. A false
positive (FP) or false alarm, which refers to the detection of
benign traffic as an attack. A  false negative (FN) refers to
detecting attack traffic as benign traffic. A key goal of an
intrusion detection system is to minimize both the FP rate and FN
rate. We use the following parameters to evaluate the proposed
model's performance:

\newcounter{5}
\begin{list}{-}
{\usecounter{5}} \item True Positive (TP): the number of
observations correctly assigned to the positive class.\vspace{1
mm}

\item False Positive (FP): the number of observations assigned by
the model to the positive class.\vspace{1 mm}

\item True Positive Rate (TPR) reflects the classifier's ability
to detect members of the positive class \[ TPR = \frac{TP} {(TP +
FN)} \]\vspace{1 mm}

\item False Positive Rate (FPR) reflects the frequency with which
the classifier makes a mistake by classifying normal state as
pathological \[ FPR = \frac {FP}{(FP + TN)} \]
\end{list}
An ROC curve plots TPR versus FPR at different classification
thresholds. Lowering the classification threshold classifies more
items as positive, and thus, increasing both false positives and
true positives.

\begin{figure}[hbt!]
    \includegraphics[width=9cm, height=6cm]{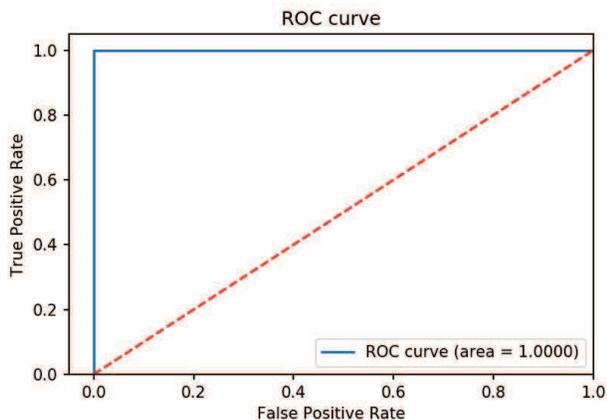}
    \caption{ROC Curve of the Proposed Model}
\end{figure}
As defining normality with a descriptive feature set is difficult,
anomalies raised by systems can sometime be detected with false
alarms (false positives) or missed alerts (false negatives). With
the ROC curve, the closer the graph is to the top and left-hand
borders, the more accurate the test. Similarly, the closer the
graph to the diagonal, the less accurate the test. The
experimental results obtained on the proposed model are as
follows:

Precision: 0.9979

Recall: 1.00

Number of True Positive: 1097

Number of Samples: 1097

True Positive Rate: 1.00

Number of False Positive: 7

Number of samples: 2200

False Positive Rate: 0.0032

\section{Conclusion}
We discussed an intrusion detection model using deep learning. The
proposed model detects web application attacks autonomously in
real-time. The model uses auto-encoder that can learn from the
sequences of word and weight each word or character according to
them. The experimental results show that the proposed model can
detect web applications attack with low false positive rate and
true positive rate is 1. Because of less volume of labeled
categorized anomalous dataset, the proposed classification engine
is not 100 percent accurate; however, the classification can be
improved with optimized training with a large volume of dataset,
which is left as the future scope of the work.

\end{document}